\documentclass[12pt,epsf,fleqn]{article}
\usepackage{epsfig}
\usepackage[dvips]{color}
\begin{document}
\title{Neutral Higgs bosons in the MNMSSM with explicit CP violation}
\author{S. W. Ham$^{(1,2)}$, J. O. Im$^{(3)}$, and S. K. Oh$^{(2,3)}$
\\
\\
{\it $^{\rm (1)}$ Department of Physics, KAIST, Daejeon 305-701, Korea} \\
{\it $^{\rm (2)}$ Center for High Energy Physics, Kyungpook National University}\\
{\it  Daegu 702-701, Korea} \\
{\it $^{\rm (3)}$ Department of Physics, Konkuk University, Seoul 143-701, Korea} \\
\\
\\
}
\date{}
\maketitle
\begin{abstract}
Within the framework of the minimal non-minimal supersymmetric standard model (MNMSSM) with tadpole terms,
CP violation effects in the Higgs sector are investigated at the one-loop level,
where the radiative corrections from the loops of the quark and squarks of the third generation are
taken into account.
Assuming that the squark masses are not degenerate, the radiative corrections
due to the stop and sbottom quarks give rise to CP phases,
which trigger the CP violation explicitly in the Higgs sector of the MNMSSM.
The masses, the branching ratios for dominant decay channels,
and the total decay widths of the five neutral Higgs bosons in the MNMSSM are calculated
in the presence of the explicit CP violation.
The dependence of these quantities on the CP phases is quite recognizable, for given parameter values.
\end{abstract}
\vfil
\eject

\section{INTRODUCTION}

Any supersymmetric standard models should have broken supersymmetry (SUSY)
in order to be phenomenologically realistic.
An intensively studied method of breaking the SUSY is the inclusion of soft breaking terms
in the Lagrangian density of the model [1].
Phenomenological analyses of the supersymmetric standard model with soft breaking terms might become
more realistic as well as more interesting if the soft breaking terms contain or induce some complex phases,
since the presence of the complex phases may give rise to CP mixing in the Higgs sector of the model [2].

The simplest version of the supersymmetric standard model is the minimal supersymmetric standard model (MSSM),
which has just two Higgs doublets in its Higgs sector.
The MSSM at the tree level cannot accommodate explicit nor spontaneous CP violation,
because any complex phases in the Higgs sector of the MSSM can always be eliminated by rotating the Higgs fields.
Even at the one-loop level, spontaneous CP violation is disfavored by the MSSM
because a very light neutral Higgs boson is required, which has already been ruled out by experiments [3].
On the other hand, it is known that explicit CP violation is viable in the MSSM at the one-loop level,
since the radiative corrections due to quarks and squarks yield
the CP mixings between the neutral Higgs bosons [4-7].

A number of nonminimal versions of the supersymmetric standard model have been introduced in the literature,
by including a Higgs singlet to the Higgs sector of the MSSM [8],
such as the next-to-minimal supersymmetric standard model (NMSSM) with $Z_3$ symmetry [9-16],
the general NMSSM with  broken $Z_3$ [17],
the minimal non-minimal supersymmetric standard model (MNMSSM) with tadpole terms [18],
and the minimal supersymmetric models with an additional $U(1)$ [19,20], to cite a few of them.
An advantage of them over the MSSM is that the vacuum expectation value of the Higgs singlet may
dynamically solve the problem of dimensional $\mu$-parameter in the MSSM [21].

The NMSSM, which is most popular among those nonminimal versions of the supersymmetric model,
has been extensively studied in various aspects of phenomenology, including explicit CP violation [13-16].
In the NMSSM, explicit CP violation may take place at the tree level,
since the NMSSM may have one nontrivial CP phase after redefining the Higgs fields.
By assuming the degeneracy of the stop quark masses in the Higgs sector of the NMSSM,
it is found that large explicit CP violation may be realized
as the vacuum expectation value of the neutral Higgs singlet
in the NMSSM approaches to the electroweak scale [13].
We have elsewhere studied the effects of explicit CP violation
in the NMSSM on the neutral and charged Higgs boson masses at the one-loop level
by considering radiative corrections due to various particles and their superpartners [15].
Also, we have recently calculated the Higgs decays
within the context of the NMSSM with explicit CP violation [16].

The MNMSSM is different from the NMSSM in the sense that explicit CP violation
is not possible at the tree level in the MNMSSM.
This is because the Higgs sector of the MNMSSM, unlike the NMSSM, can always absorb
the CP-violating complex phase by rotating the relevant Higgs fields such that its tree-level Higgs potential
cannot have any CP phase.
Thus, within the context of explicit CP violation in the Higgs sector,
the MNMSSM is similar to the MSSM with an additional $U(1)$ rather than the NMSSM.
At the tree level, the MSSM with an additional $U(1)$ can always eliminate
any complex CP phase in its Higgs sector by employing the tadpole minimum condition or
by rotating the three neutral Higgs fields [20].
The CP symmetry of the MNMSSM may  be violated at the one-loop level in an explicit way
as radiative corrections are taken into account.

In this article, we study the explicit CP violation scenario at the one-loop level
in the Higgs sector of the MNMSSM.
We calculate the radiatively corrected masses of the neutral Higgs bosons in the MNMSSM.
The dominant decay modes for the neutral Higgs bosons into heavy fermion pairs, gluon pairs,
and weak boson pairs, are studied.
We are interested in the dependency of those decays on the CP phases
in the explicit CP violation scenario at the one-loop level.
The branching ratios for the neutral Higgs bosons with CP-undefined states are found to vary
significantly against the CP phases, arising from the squark masses of the third generation.

\section{THE CP-VIOLATING HIGGS POTENTIAL}

The Higgs sector of the MNMSSM consists of two Higgs doublet superfields
${\widehat H}_1 = ({\widehat H}_1^0, {\widehat H}_1^-)$,
${\widehat H}_2 = ({\widehat H}_2^+, {\widehat H}_2^0)$, and a Higgs singlet superfield ${\widehat N}$.
Keeping only the Yukawa couplings for the third generation of quarks, the superpotential of MNMSSM
may be written as
\begin{eqnarray}
W = h_t \varepsilon_{ij} {\widehat Q}^i {\widehat t}_R^c {\widehat H}_2^j
- h_b \varepsilon_{ij} {\widehat Q}^i {\widehat b}_R^c {\widehat H}_1^j
- h_{\tau} \varepsilon_{ij} {\widehat L}^i {\widehat \tau}_R^c {\widehat H}_1^j
+ \lambda \varepsilon_{ij} {\widehat H}_1^i {\widehat H}_2^j {\widehat N}
\end{eqnarray}
where $\varepsilon_{ij}$ is totally antisymmetric with $\varepsilon_{12}=-\varepsilon_{21}=1$,
${\widehat Q}$ and ${\widehat L}$ are the SU(2) doublet quark and lepton superfields of the third
generation, respectively, ${\widehat t}_R^c$, ${\widehat b}_R^c$ and ${\widehat \tau}_R^c$ are the
SU(2) singlet top, bottom, and tau superfields respectively, $h_t$,
$h_b$, and $h_{\tau}$ are the Yukawa coupling coefficients of top,
bottom, and tau superfields, respectively, and $\lambda$ is a dimensionless coupling coefficient.

The tree-level Higgs potential, $V^0$, of the MNMSSM may be decomposed into $D$-terms, $F$-terms,
the soft terms, and the tadpole terms as
\[
    V^0 = V_D + V_F + V_{\rm S} + V_{\rm T} \ ,
\]
where
\begin{eqnarray}
V_D & = & {g_2^2 \over 8} (H_1^{\dag} \vec \sigma H_1 + H_2^{\dag} \vec \sigma H_2)^2
+ {g_1^2\over 8}(|H_2|^2 - |H_1|^2)^2  \ , \cr
V_F & = & |\lambda|^2[(|H_1|^2+|H_2|^2)|N|^2+|\varepsilon_{ij} H_1^i H_2^j|^2] \ , \cr
V_{\rm S} & = & m_{H_1}^2|H_1|^2 + m_{H_2}^2|H_2|^2 + m_N^2|N|^2
- (\lambda A_{\lambda} \varepsilon_{ij} H_1^i H_2^j N + {\rm H.c.} ) \ , \cr
V_{\rm T} & = &\mbox{} - (\xi^3 N + {\rm H.c.} )   \ ,
\end{eqnarray}
with $g_1$ and $g_2$ being the $U(1)$ and $SU(2)$ gauge coupling constants, respectively,
$\vec \sigma$ being the Pauli matrices,
$A_{\lambda}$ being the trilinear soft SUSY breaking parameter with mass dimension,
$m_{H_1}$, $m_{H_2}$, and $m_N$ are the soft SUSY breaking masses,
and $\xi$ is the tadpole coefficient.

Note that the above tree-level Higgs potential would additionally have a global $U(1)$ Peccei-Quinn symmetry
if there is no tadpole term.
The  global $U(1)$ Peccei-Quinn symmetry gives a natural solution to the strong CP problem [22].
However, it leads to the existence of a massless pseudo-Goldstone boson
which emerges from the tree-level Higgs potential
since the determinant of the pseudoscalar Higgs boson mass matrix is zero [23].
This is the Weinberg-Wilczek axion.
The  global $U(1)$ Peccei-Quinn symmetry is eventually broken by the quantum effects arising
from the triangle anomaly,
since the Noether current for the global $U(1)$ transformation is anomalous, and the
Weinberg-Wilczek axion acquires a small mass due to instanton effects.
Its mass is inversely proportional to the scale of the axion decay constant $f_A$ as
\[
m_A = 0.6 {\rm eV} {10^7 \over f_A} \ ,
\]
where $f_A$ is assumed to be the electroweak scale.
By considering the Weinberg-Wilczek axion as the candidate for the cold dark matter, cosmology
estimates $m_A \sim 10^{-3} - 10^{-6} {\rm eV}$ [24].
This information predicts that the scale of the axion decay constant is bounded as  $f_A > 6 \times 10^8$ GeV.

The Weinberg-Wilczek axion has been excluded by the negative results of experimental searches.
On the other hand, in such models as the Kim-Shifman-Vainshtein-Zakharov model [25] or
the Dine-Fischler-Srednicki-Zhitnitskii model [26], the scale of the axion decay constant
can be very high and thus the axion in these models becomes practically invisible.
Therefore, it may survive without contradicting the experimental constraints.

In order to avoid the Weinberg-Wilczek axion in the MNMSSM,
we introduce the tadpole term in the Higgs potential, as seen in Eq.(2),
Thus, the global $U(1)$ Peccei-Quinn symmetry in the MNMSSM is explicitly broken.
The tadpole coefficient $\xi$ in the tadpole term is intrinsically a free parameter.
However, it might be assumed that it is the same order of the SUSY breaking scale
from a phenomenological point of view.

We would like to show that the Higgs sector of the MNMSSM cannot invoke explicit CP violation at the tree level.
Note that, if the above tree-level Higgs potential contain any complex phases,
the possibility would be that $A_{\lambda}$, $\lambda$, or $\xi$ are complex.
Among them, without loss of generality, $A_{\lambda}$ may be assumed to be real,
since its phase can always be absorbed into the phase of $\lambda$.
Further, $\lambda$ can be made real by adjusting the phases of the Higgs doublets,
while $\xi$ can also be made real by adjusting the phase of the Higgs singlet.
Consequently, the Higgs potential of the MNMSSM at the tree level can only have real parameters,
hence no explicit CP violation.
It is also impossible for the Higgs sector of the MNMSSM at the tree level to invoke spontaneous CP violation.

In terms of the physical Higgs fields, the Higgs doublets and the Higgs singlet may generally be expressed as
\begin{eqnarray}
\begin{array}{lll}
        H_1 & = & \left ( \begin{array}{c}
          v_1 + S_1 + i \sin \beta P_1   \cr-
          \sin \beta C^{+ *}
  \end{array} \right )  \ ,  \cr
        H_2 & = & \left ( \begin{array}{c}
          \cos \beta C^+           \cr
          (v_2 + S_2 + i \cos \beta P_1) e^{i \theta}
  \end{array} \right )   \ ,  \cr
        N & = & \left ( \begin{array}{c}
          x + S_3 + i P_2
  \end{array} \right ) e^{i \delta}   \ ,
\end{array}
\end{eqnarray}
where $S_i$ ($i = 1,2,3$) are the neutral scalar Higgs fields, $P_i$ ($i=1, 2$) are
the neutral pseudoscalar Higgs fields, $C^+$ is the charged Higgs field, and $v_1$, $v_2$, and $x$ are respectively
the vacuum expectation values of the neutral Higgs fields with $v = \sqrt{v_1^2 + v_2^2}$ = 174 GeV
and $\tan \beta = v_2/v_1$,
$\theta$ is the relative phase between $H_1$ and $H_2$, and $\delta$ is the phase of $N$.

Now, the tree-level tadpole minimum conditions upon the tree-level Higgs potential of the MNMSSM,
which are derived from the first derivatives of the tree-level Higgs potential
with respect to the two pseudoscalar Higgs fields, may be written as
\begin{eqnarray}
0 & = & 2 \lambda v x A_{\lambda} \sin \theta_1 \ , \cr
0 & = & \lambda v^2 A_{\lambda} \sin 2 \beta \sin \theta_1 + 2 \xi^3 \sin \theta_2 \ ,
\end{eqnarray}
where the two phases $\theta_1$ and $\theta_2$ are defined
as $\theta_1 = \theta + \delta$ and $\theta_2 = \delta$.
It is straightforward that the tadpole minimum conditions are satisfied only
when both $\theta_1$ and $\theta_2$ are zero, in other words, only if $\theta = \delta = 0$.
Therefore, explicit CP violation is not possible in the Higgs sector of the MNMSSM at the tree level.

In order to accommodate any complex phases so as to induce the CP violation
between the scalar and pseudoscalar Higgs bosons in the present model, one has to consider higher-order corrections.
The radiative corrections due to top and stop quarks are known
to affect significantly the tree-level Higgs sector of supersymmetric models.
The contribution of the bottom and sbottom quark loops is not negligible for very large $\tan \beta$.

The Higgs potential at the one-loop level may be written as
\[
    V = V^0 + V^1
\]
where $V^1$ is the one-loop effective Higgs potential including the radiative corrections
due to quarks and squarks of the third generation, which is explicitly given as [27,18]
\begin{equation}
V^1 = \sum_{i = 1}^2 {3 {\cal M}_{{\tilde q}_i}^4 \over 32 \pi^2}
  \left (\log {{\cal M}_{{\tilde q}_i}^2 \over \Lambda^2} - {3\over 2} \right )
  - {3 {\cal M}_q^4 \over 16 \pi^2} \left (\log {{\cal M}_q^2 \over \Lambda^2}
  - {3\over 2} \right ) \ ,
\end{equation}
where ${\cal M}_q$ ($q=t,b$) and ${\cal M}_{{\tilde q}_i}$ ($i$ = 1 ,2) are respectively
the quark and squark masses given as functions of the Higgs fields, and
$\Lambda$ is the renormalization scale in the modified minimal subtraction scheme.

After the spontaneous breakdown of the electroweak symmetry,
the quark masses of the third generation are given by $m_t = h_t v_2$ and $m_b = h_b v_1$,
and the squark masses of the third generation are given by
\begin{eqnarray}
m_{{\tilde t}_1, {\tilde t}_2}^2 & = & m_Q^2 + m_t^2 \mp h_t
\sqrt{A_t^2 v_2^2 + \lambda^2 v_1^2 x^2 + 2 \lambda A_t v_1 v_2 x \cos \phi_t} \ , \cr
m_{{\tilde b}_1, {\tilde b}_2}^2 & = & m_Q^2 + m_b^2 \mp m_b
\sqrt{A_b^2 + \lambda^2 x^2 \tan^2 \beta + 2 \lambda A_b x \tan \beta \cos \phi_b} \ ,
\end{eqnarray}
where $m_Q$ is the soft SUSY breaking mass,
$A_t$ and $A_b$ are the trilinear SUSY breaking parameters with mass dimension,
and two phases $\phi_t$ and $\phi_b$ are given as
\begin{eqnarray}
&& \phi_t = \phi_{A_t} + \theta + \delta \ ,  \cr
&& \phi_b = \phi_{A_b} + \theta + \delta \ , \nonumber
\end{eqnarray}
where  $\phi_{A_t}$ and $\phi_{A_b}$ are respectively the phases of $A_t$ and $A_b$,
which are in general assumed to be complex.
In the expressions for the squark masses, $A_t$ and $A_b$ are now real parameters
because their phases are already taken out.
Note that the $D$-terms are not included in the squark masses.

Now, at the one-loop level, the tadpole minimum conditions may be written as
\begin{eqnarray}
0 & = & A_{\lambda} \sin \theta_1 + {3 h_t^2 \over 16 \pi^2 } A_t \sin \phi_t
f(m_{{\tilde t}_1}^2, m_{{\tilde t}_2}^2)
+ {3 h_b^2 \over 16 \pi^2} A_b \sin \phi_b f(m_{{\tilde b}_1}^2, m_{{\tilde b}_2}^2) \ , \cr
0 & = & A_{\lambda} \lambda v^2 \sin 2 \beta \sin \theta_1 + 2 \xi^3 \sin \theta_2
+ {3 h_t^2 \over 16 \pi^2 } A_t \lambda v^2 \sin 2 \beta \sin \phi_t
f(m_{{\tilde t}_1}^2, m_{{\tilde t}_2}^2) \cr
& &\mbox{}+ {3 h_b^2 \over 16 \pi^2} A_b \lambda v^2 \sin 2 \beta \sin \phi_b
f(m_{{\tilde b}_1}^2, m_{{\tilde b}_2}^2) \ ,
\end{eqnarray}
where  the scale-dependent function $f(m_x^2, m_y^2)$ is defined as
\begin{equation}
 f(m_x^2, m_y^2) = {1 \over (m_y^2 - m_x^2)} \left[  m_x^2 \log {m_x^2 \over \Lambda^2} - m_y^2
\log {m_y^2 \over \Lambda^2} \right] + 1 \ .
\end{equation}

It is quite obvious that  $\theta_1 = 0$ does not satisfy the first tadpole minimum condition
at the one-loop level.
Unlike the tree-level case, $\theta_1$ at the one-loop level is not zero but dependent on the other parameters.
Let us rename $\theta_1$ at the one-loop level as $\phi_0$ hereafter, in order to avoid any confusion
with the tree-level $\theta_1$ that is zero.
On the other hand, one can easily see that $\sin\theta_2 = 0$ satisfies the second tadpole minimum condition,
by substituting the first tadpole minimum condition into the second one.
In other words, we have $\theta_2 = \delta = 0$ and $\theta_1 = \theta$, renamed as $\phi_0$,
which is expressed in terms of the other parameters.
Thus, we are left with $\phi_t$ and $\phi_b$ at the one-loop level.
Consequently, the Higgs sector of the MNMSSM may eventually have
two physical CP phase $\phi_t$ and $\phi_b$ at the one-loop level
even if there is no complex phase at the tree level.

The $5\times 5$ symmetric mass matrix, $M$, for the five neutral Higgs bosons
at the one-loop level in the MNMSSM is obtained from
the second derivatives of $V$ with respect to $S_1$, $S_2$, $S_3$, $P_1$ and $P_2$.
In the basis of ($S_1, S_2, P_1, S_3, P_2$), where we permute $P_1$ and $S_3$ for convenience,
the matrix elements of $M$ are obtained as
\begin{eqnarray}
M_{11} & = & M_{11}^t + M_{11}^b + (m_Z \cos \beta)^2 + m_A^2 \sin^2 \beta , \cr
M_{22} & = & M_{22}^t + M_{22}^b + (m_Z \sin \beta)^2 + m_A^2 \cos^2 \beta , \cr
M_{33} & = & M_{33}^t + M_{33}^b + m_A^2  \ , \cr
M_{44} & = & M_{44}^t + M_{44}^b + {v^2 \over 4 x^2} m_A^2 \sin^2 2 \beta  + {\xi^3 \over x} \ , \cr
M_{55} & = & M_{55}^t + M_{55}^b + {v^2 \over 4 x^2} m_A^2 \sin^2 2 \beta  + {\xi^3 \over x} \ , \cr
M_{12} & = & M_{12}^t + M_{12}^b + (2 \lambda^2 v^2 - m_Z^2 - m_A^2) \sin \beta \cos \beta \ , \cr
M_{13} & = & M_{13}^t + M_{13}^b  \ , \cr
M_{14} & = & M_{14}^t + M_{14}^b - {v \over x} m_A^2 \sin^2 \beta \cos \beta + 2 v \lambda^2 x \cos \beta \ , \cr
M_{15} & = & M_{15}^t + M_{15}^b  \ , \cr
M_{23} & = & M_{23}^t + M_{23}^b  \ , \cr
M_{24} & = & M_{24}^t + M_{24}^b - {v \over x} m_A^2 \sin \beta \cos^2 \beta + 2 v \lambda^2 x \sin \beta \ , \cr
M_{25} & = & M_{25}^t + M_{25}^b  \ , \cr
M_{34} & = & M_{34}^t + M_{34}^b  \ , \cr
M_{35} & = & M_{35}^t + M_{35}^b + {v \over x} m_A^2 \sin \beta \cos \beta  \ , \cr
M_{45} & = & M_{45}^t + M_{45}^b  \ ,
\end{eqnarray}
where $M^t_{ij}$ ($i,j$ =1-5) are the radiative corrections due to top and stop quarks,
$M^b_{ij}$ ($i,j$ =1-5) are the radiative corrections due to bottom and sbottom quarks,
$m_Z^2 = (g_1^2 + g_2^2) v^2/2$ is the squared mass of the neutral gauge boson,
and $m_A^2$ is introduced for convenience as
\begin{eqnarray}
m_A^2 & = & {\lambda x A_{\lambda} \cos \phi_0 \over \sin \beta \cos \beta}
+ {3 m_t^2 A_t \lambda x \cos \phi_t \over 16 \pi^2 v^2 \sin^3 \beta \cos \beta}
f (m_{{\tilde t}_1}^2, m_{{\tilde t}_2}^2) \cr
& &\mbox{} + {3 m_b^2 A_b \lambda x \cos \phi_b \over 16 \pi^2 v^2 \cos^3 \beta \sin \beta}
f (m_{{\tilde b}_1}^2, m_{{\tilde b}_2}^2) \ .
\end{eqnarray}
Explicitly, $M^t_{ij}$ are given as
\begin{eqnarray}
M_{11}^t & = & {3 m_t^4 \lambda^2 x^2 \Delta_{{\tilde t}_1}^2 \over 8 \pi^2  v^2 \sin^2 \beta}
{g(m_{\tilde{t}_1}^2, \ m_{\tilde{t}_2}^2) \over (m_{\tilde{t}_2}^2 - m_{\tilde{t}_1}^2)^2}   \ , \cr
M_{22}^t & = & {3 m_t^4 A_t^2 \Delta_{{\tilde t}_2}^2 \over 8 \pi^2  v^2 \sin^2 \beta}
{g(m_{\tilde{t}_1}^2, \ m_{\tilde{t}_2}^2) \over (m_{\tilde{t}_2}^2 - m_{\tilde{t}_1}^2)^2}
+ {3 m_t^4 A_t \Delta_{{\tilde t}_2} \over 4 \pi^2 v^2 \sin^2 \beta}
{\log (m_{\tilde{t}_2}^2 / m_{\tilde{t}_1}^2)  \over (m_{\tilde{t}_2}^2 - m_{\tilde{t}_1}^2)} \cr
& &\mbox{} + {3 m_t^4 \over 8 \pi^2 v^2 \sin^2 \beta}
\log \left ( {m_{\tilde{t}_1}^2  m_{\tilde{t}_2}^2 \over m_t^4} \right ) \ , \cr
M_{33}^t & = & {3 m_t^4 \lambda^2 x^2 A_t^2 \sin^2 \phi_t \over 8 \pi^2 v^2 \sin^4 \beta}
{g(m_{\tilde{t}_1}^2, \ m_{\tilde{t}_2}^2) \over (m_{\tilde{t}_2}^2 - m_{\tilde{t}_1}^2 )^2}  \ , \cr
M_{44}^t & = & {3 m_t^4 \lambda^2 \Delta_{{\tilde t}_1}^2 \over 8 \pi^2 \tan^2 \beta}
{g(m_{\tilde{t}_1}^2, \ m_{\tilde{t}_2}^2) \over (m_{\tilde{t}_2}^2 - m_{\tilde{t}_1}^2 )^2}  \ , \cr
M_{55}^t & = & {3 m_t^4 \lambda^2 A_t^2 \sin^2 \phi_t \over 8 \pi^2 \tan^2 \beta}
{g(m_{\tilde{t}_1}^2, \ m_{\tilde{t}_2}^2) \over (m_{\tilde{t}_2}^2 - m_{\tilde{t}_1}^2 )^2}  \ ,  \cr
M_{12}^t & = & {3 m_t^4 \lambda x A_t \Delta_{{\tilde t}_1} \Delta_{{\tilde t}_2} \over 8 \pi^2 v^2 \sin^2 \beta}
{g(m_{\tilde{t}_1}^2, \ m_{\tilde{t}_2}^2) \over (m_{\tilde{t}_2}^2 - m_{\tilde{t}_1}^2)^2}
+ {3 m_t^4 \lambda x \Delta_{{\tilde t}_1} \over 8 \pi^2 v^2 \sin^2 \beta}
{\log (m_{\tilde{t}_2}^2 / m_{\tilde{t}_1}^2) \over (m_{\tilde{t}_2}^2 - m_{\tilde{t}_1}^2)}  \ , \cr
M_{13}^t & = & \mbox{} - {3 m_t^4 \lambda^2 x^2 A_t \Delta_{{\tilde t}_1} \sin \phi_t \over 8 \pi^2 v^2 \sin^3 \beta}
{g(m_{\tilde{t}_1}^2, \ m_{\tilde{t}_2}^2) \over (m_{\tilde{t}_2}^2 - m_{\tilde{t}_1}^2)^2 } \ , \cr
M_{14}^t & = & {3 m_t^4 \lambda^2 x \Delta_{{\tilde t}_1}^2 \over 8 \pi^2 v \sin \beta \tan \beta}
{g(m_{\tilde{t}_1}^2, \ m_{\tilde{t}_2}^2) \over (m_{\tilde{t}_2}^2 - m_{\tilde{t}_1}^2)^2 }
- {3 m_t^2 \lambda^2 x \cot \beta \over 8 \pi^2 v \sin \beta} f(m_{\tilde{t}_1}^2, \ m_{\tilde{t}_2}^2)  , \cr
M_{15}^t & = &\mbox{} - {3 m_t^4 \lambda^2 x A_t \Delta_{{\tilde t}_1} \sin \phi_t \over 8 \pi^2 v \sin \beta \tan \beta}
{g(m_{\tilde{t}_1}^2, \ m_{\tilde{t}_2}^2) \over (m_{\tilde{t}_2}^2 - m_{\tilde{t}_1}^2)^2 } \ , \cr
M_{23}^t & = & \mbox{} - {3 m_t^4 \lambda x A_t^2 \Delta_{{\tilde t}_2} \sin \phi_t \over 8 \pi^2 v^2 \sin^3 \beta}
{g(m_{\tilde{t}_1}^2, \ m_{\tilde{t}_2}^2) \over (m_{\tilde{t}_2}^2 - m_{\tilde{t}_1}^2)^2 }
- {3 m_t^4 \lambda x A_t \sin \phi _t \over 8 \pi^2 v^2 \sin^3 \beta}
{\log (m_{\tilde{t}_2}^2 / m_{\tilde{t}_1}^2) \over (m_{\tilde{t}_2}^2 - m_{\tilde{t}_1}^2)}  \ , \cr
M_{24}^t & = & {3 m_t^4 \lambda A_t \Delta_{{\tilde t}_1} \Delta_{{\tilde t}_2} \over 8 \pi^2 v \sin \beta \tan \beta}
{g(m_{\tilde{t}_1}^2, \ m_{\tilde{t}_2}^2) \over (m_{\tilde{t}_2}^2 - m_{\tilde{t}_1}^2)^2}
 + {3 m_t^4 \lambda \Delta_{{\tilde t}_1} \over 8 \pi^2 v \sin \beta \tan \beta}
{\log (m_{\tilde{t}_2}^2 / m_{\tilde{t}_1}^2) \over (m_{\tilde{t}_2}^2 - m_{\tilde{t}_1}^2) }  , \cr
M_{25}^t & = & \mbox{} - {3 m_t^4 \lambda A_t^2 \Delta_{{\tilde t}_2} \sin \phi_t \over 8 \pi^2 v \sin \beta \tan \beta}
{g(m_{\tilde{t}_1}^2, \ m_{\tilde{t}_2}^2) \over (m_{\tilde{t}_2}^2 - m_{\tilde{t}_1}^2)^2 }
- {3 m_t^4 \lambda A_t \sin \phi_t \over 8 \pi^2 v \sin \beta \tan \beta}
{ \log (m_{\tilde{t}_2}^2 / m_{\tilde{t}_1}^2) \over (m_{\tilde{t}_2}^2 - m_{\tilde{t}_1}^2)}  \ , \cr
M_{34}^t & = & \mbox{} - {3 m_t^4 \lambda^2 x A_t \Delta_{{\tilde t}_1} \sin \phi_t \over 8 \pi^2 v \sin^2 \beta \tan \beta}
{g(m_{\tilde{t}_1}^2, \ m_{\tilde{t}_2}^2) \over (m_{\tilde{t}_2}^2 - m_{\tilde{t}_1}^2)^2 } \ , \cr
M_{35}^t & = & \mbox{} {3 m_t^4 \lambda^2 x A_t^2 \sin^2 \phi_t \over 8 \pi^2 v \sin^2 \beta \tan \beta}
{g(m_{\tilde{t}_1}^2, \ m_{\tilde{t}_2}^2) \over (m_{\tilde{t}_2}^2 - m_{\tilde{t}_1}^2)^2 }  \ , \cr
M_{45}^t & = & \mbox{} - {3 m_t^4 \lambda^2 A_t \Delta_{{\tilde t}_1} \sin \phi_t \over 8 \pi^2 \tan^2 \beta}
{ g(m_{\tilde{t}_1}^2, \ m_{\tilde{t}_2}^2) \over (m_{\tilde{t}_2}^2 - m_{\tilde{t}_1}^2)^2 } \ ,
\end{eqnarray}
and $M_{ij}^b$ are given as
\begin{eqnarray}
M_{11}^b & = & {3 m_b^4 A_b^2 \Delta_{{\tilde b}_1}^2 \over 8 \pi^2  v^2 \cos^2 \beta}
{g(m_{\tilde{b}_1}^2, \ m_{\tilde{b}_2}^2) \over (m_{\tilde{b}_2}^2 - m_{\tilde{b}_1}^2)^2}
+ {3 m_b^4 A_b \Delta_{{\tilde b}_1} \over 4 \pi^2 v^2 \cos^2 \beta}
{\log (m_{\tilde{b}_2}^2 / m_{\tilde{b}_1}^2)  \over (m_{\tilde{b}_2}^2 - m_{\tilde{b}_1}^2)} \cr
& &\mbox{} + {3 m_b^4 \over 8 \pi^2 v^2 \cos^2 \beta}
\log \left ( {m_{\tilde{b}_1}^2  m_{\tilde{b}_2}^2 \over m_b^4} \right ) \ , \cr
M_{22}^b & = & {3 m_b^4 \lambda^2 x^2 \Delta_{{\tilde b}_2}^2 \over 8 \pi^2  v^2 \cos^2 \beta}
{g(m_{\tilde{b}_1}^2, \ m_{\tilde{b}_2}^2) \over (m_{\tilde{b}_2}^2 - m_{\tilde{b}_1}^2)^2}   \ , \cr
M_{33}^b & = & {3 m_b^4 \lambda^2 x^2 A_b^2 \sin^2 \phi_b \over 8 \pi^2 v^2 \cos^4 \beta}
{g(m_{\tilde{b}_1}^2, \ m_{\tilde{b}_2}^2) \over (m_{\tilde{b}_2}^2 - m_{\tilde{b}_1}^2 )^2} \ , \cr
M_{44}^b & = & {3 m_b^4 \lambda^2 \Delta_{{\tilde b}_2}^2 \over 8 \pi^2 \cot^2 \beta}
{g(m_{\tilde{b}_1}^2, \ m_{\tilde{b}_2}^2) \over (m_{\tilde{b}_2}^2 - m_{\tilde{b}_1}^2 )^2}  \ , \cr
M_{55}^b & = & {3 m_b^4 \lambda^2 A_b^2 \sin^2 \phi_b \over 8 \pi^2 \cot^2 \beta}
{g(m_{\tilde{b}_1}^2, \ m_{\tilde{b}_2}^2) \over (m_{\tilde{b}_2}^2 - m_{\tilde{b}_1}^2 )^2} \ ,  \cr
M_{12}^b & = & {3 m_b^4 \lambda x A_b \Delta_{{\tilde b}_1} \Delta_{{\tilde b}_2} \over 8 \pi^2 v^2 \cos^2 \beta}
{g(m_{{\tilde b}_1}^2, \ m_{{\tilde b}_2}^2) \over (m_{{\tilde b}_2}^2 - m_{{\tilde b}_1}^2)^2}
+ {3 m_b^4 \lambda x \Delta_{{\tilde b}_2} \over 8 \pi^2 v^2 \cos^2 \beta}
{\log (m_{{\tilde b}_2}^2 / m_{{\tilde b}_1}^2)  \over (m_{{\tilde b}_2}^2 - m_{{\tilde b}_1}^2)} \ , \cr
M_{13}^b & = & \mbox{} - {3 m_b^4 \lambda x A_b^2 \Delta_{{\tilde b}_1} \sin \phi_b \over 8 \pi^2 v^2 \cos^3 \beta}
{g(m_{\tilde{b}_1}^2, \ m_{\tilde{b}_2}^2) \over (m_{\tilde{b}_2}^2 - m_{\tilde{b}_1}^2)^2 }
- {3 m_b^4 \lambda x A_b \sin \phi_b \over 8 \pi^2 v^2 \cos^3 \beta}
{\log (m_{\tilde{b}_2}^2 / m_{\tilde{b}_1}^2) \over (m_{\tilde{b}_2}^2 - m_{\tilde{b}_1}^2)}  \ , \cr
M_{14}^b & = & {3 m_b^4 \lambda A_b \Delta_{{\tilde b}_1} \Delta_{{\tilde b}_2} \over 8 \pi^2 v \cos \beta \cot \beta}
{g(m_{\tilde{b}_1}^2, \ m_{\tilde{b}_2}^2) \over (m_{\tilde{b}_2}^2 - m_{\tilde{b}_1}^2)^2}
 + {3 m_b^4 \lambda \Delta_{{\tilde b}_2} \over 8 \pi^2 v \cos \beta \cot \beta}
{\log (m_{\tilde{b}_2}^2 / m_{\tilde{b}_1}^2) \over (m_{\tilde{b}_2}^2 - m_{\tilde{b}_1}^2) }  , \cr
M_{15}^b & = & \mbox{} - {3 m_b^4 \lambda A_b^2 \Delta_{{\tilde b}_1} \sin \phi_b \over 8 \pi^2 v \cos \beta \cot \beta}
{g(m_{\tilde{b}_1}^2, \ m_{\tilde{b}_2}^2) \over (m_{\tilde{b}_2}^2 - m_{\tilde{b}_1}^2)^2 }
- {3 m_b^4 \lambda A_b \sin \phi_b \over 8 \pi^2 v \cos \beta \cot \beta}
{ \log (m_{\tilde{b}_2}^2 / m_{\tilde{b}_1}^2) \over (m_{\tilde{b}_2}^2 - m_{\tilde{b}_1}^2)}  \ , \cr
M_{23}^b & = & \mbox{} - {3 m_b^4 \lambda^2 x^2 A_b \Delta_{{\tilde b}_2} \sin \phi_b \over 8 \pi^2 v^2 \cos^3 \beta}
{g(m_{\tilde{b}_1}^2, \ m_{\tilde{b}_2}^2) \over (m_{\tilde{b}_2}^2 - m_{\tilde{b}_1}^2)^2 } \ , \cr
M_{24}^b & = & {3 m_b^4 \lambda^2 x \Delta_{{\tilde b}_2}^2 \over 8 \pi^2 v \cos \beta \cot \beta}
{g(m_{\tilde{b}_1}^2, \ m_{\tilde{b}_2}^2) \over (m_{\tilde{b}_2}^2 - m_{\tilde{b}_1}^2)^2 }
- {3 m_b^2 \lambda^2 x \tan \beta \over 8 \pi^2 v \cos \beta} f(m_{\tilde{b}_1}^2, \ m_{\tilde{b}_2}^2)  , \cr
M_{25}^b & = &\mbox{} - {3 m_b^4 \lambda^2 x A_b \Delta_{{\tilde b}_2} \sin \phi_b \over 8 \pi^2 v \cos \beta \cot \beta}
{g(m_{\tilde{b}_1}^2, \ m_{\tilde{b}_2}^2) \over (m_{\tilde{b}_2}^2 - m_{\tilde{b}_1}^2)^2 } \ , \cr
M_{34}^b & = & \mbox{} - {3 m_b^4 \lambda^2 x A_b \Delta_{{\tilde b}_2} \sin \phi_b \over 8 \pi^2 v \cos^2 \beta \cot \beta}
{g(m_{\tilde{b}_1}^2, \ m_{\tilde{b}_2}^2) \over (m_{\tilde{b}_2}^2 - m_{\tilde{b}_1}^2)^2 }   \ , \cr
M_{35}^b & = & \mbox{} {3 m_b^4 \lambda^2 x A_b^2 \sin^2 \phi_b \over 8 \pi^2 v \cos^2 \beta \cot \beta}
{g(m_{\tilde{b}_1}^2, \ m_{\tilde{b}_2}^2) \over (m_{\tilde{b}_2}^2 - m_{\tilde{b}_1}^2)^2 }  \ , \cr
M_{45}^b & = & \mbox{} - {3 m_b^4 \lambda^2 A_b \Delta_{{\tilde b}_2} \sin \phi_b \over 8 \pi^2 \cot^2 \beta}
{ g(m_{\tilde{b}_1}^2, \ m_{\tilde{b}_2}^2) \over (m_{\tilde{b}_2}^2 - m_{\tilde{b}_1}^2)^2 } \ ,
\end{eqnarray}
where
\begin{eqnarray}
 \Delta_{{\tilde t}_1} &=& A_t \cos \phi_t + \lambda x \cot \beta  \  , \cr
 \Delta_{{\tilde t}_2} & = & A_t + \lambda x \cot \beta \cos \phi_t \ , \cr
 \Delta_{{\tilde b}_1} & = & A_t + \lambda x \tan \beta \cos \phi_b \ , \cr
 \Delta_{{\tilde b}_2} &=& A_b \cos \phi_t + \lambda x \tan \beta  \  ,
\end{eqnarray}
and $g(m_x^2,m_y^2)$ is another scale-independent function that is defined as
\begin{equation}
 g(m_x^2,m_y^2) = {m_y^2 + m_x^2 \over m_x^2 - m_y^2} \log {m_y^2 \over m_x^2} + 2 \ .
\end{equation}

Among these $M_{ij}$, those that are responsible for the CP mixing
between scalar and pseudoscalar Higgs fields are $M_{13}$, $M_{23}$, $M_{15}$, $M_{25}$, $M_{34}$, and $M_{45}$
in the ($S_1, S_2, P_1, S_3, P_2$) basis.
They are clearly zero at the tree level.
Thus, there is no CP mixing between scalar and pseudoscalar neutral Higgs bosons at the tree level.
They become non-zero as the radiative corrections $M^t_{ij}$ and $M^b_{ij}$ are taken into account.
Moreover, they receive complex phases $\phi_t$ and $\phi_b$ from $M^t_{ij}$ and $M^b_{ij}$, respectively,
for the CP mixing.
Notice that $M^t_{13}$, $M^t_{23}$, $M^t_{15}$, $M^t_{25}$, $M^t_{34}$, and $M^t_{45}$ depend on $\phi_t$
while $M^b_{13}$, $M^b_{23}$, $M^b_{15}$, $M^b_{25}$, $M^b_{34}$, and $M^b_{45}$ depend on $\phi_b$.
Therefore, the magnitude of the CP mixing between scalar and pseudoscalar neutral Higgs bosons is
directly dependent on $\phi_t$ and $\phi_b$.
The CP mixing would be maximal when $\sin \phi_t = \sin \phi_b = 1$.

The five eigenvalues of $M$, denoted as $m^2_{h_i}$ ($i$ = 1-5),
define the five physical neutral Higgs bosons $h_i$ ($i$ = 1-5) as the mass eigenstates and their squared masses.
Unless $\phi_t$ or $\phi_b$ vanishes, $h_i$ ($i$ = 1-5) would not have definite CP parities.
They are in general given as the mixtures of $S_1$, $S_2$, $S_3$, $P_1$, and $P_2$.
We sort these five neutral Higgs bosons in the increasing order of their masses
such that $m^2_{h_1}$ is the smallest eigenvalue and $h_1$ is the lightest neutral Higgs boson.
If $\phi_t = \phi_b = 0$, these five neutral Higgs bosons may be classified
into three scalar and two pseudoscalar Higgs bosons.

\section{HIGGS DECAYS}

In the standard model (SM), it is well known that particles acquire their masses through the interactions
with the Higgs field.
This implies the existence of the yet-undiscovered SM Higgs boson.
The lower bound on the SM Higgs boson mass of about 114.5 GeV is determined
by means of the negative result for the Higgs searches at LEP2 experiments.
The SM does not predict its mass.
But the SM provides the decay modes as well as production rates for each possible mass range.

The decay signature of the SM Higgs boson depends on its mass.
If the Higgs boson mass is in the range of 80-140 GeV,
the CMS [28] and ATLAS [29] collaborations expect that its decay into a pair of photons would be
the most interesting channel.
In this range, other important channels are the decays into pairs of bottom quarks, charmed quarks,
and tau leptons, as well as a gluon pair.
In the mass range between 114.5 GeV and 2 $m_Z$ (=180 GeV), its decays into pairs of weak gauge bosons,
where one of the gauge boson in the pair is virtual, become dominant besides the decay channel
into a bottom quark pair.
For the SM Higgs boson mass between 2$m_Z$ and 2$m_t$ (= 350 GeV), it would almost exclusively decay
into the weak gauge boson pairs ($WW$, $ZZ$).
Its mass being larger than 2$m_t$, the SM Higgs boson would decay mainly into pairs of weak gauge bosons
or top quarks.

From the extrapolations of the relevant couplings through the Higgs decay modes in our model,
the experimental constraint on the mass of the lightest neutral Higgs boson in our model might be derived.
Therefore, the calculation of the Higgs decay modes is essential for the Higgs searches.
The total decay width of the $j$th neutral Higgs boson in our model may be assumed as [30]
\begin{eqnarray}
\Gamma(h_j) & = & \Gamma (h_j \to b b) + \Gamma (h_j \to \tau \tau)
+ \Gamma (h_j \to \mu \mu) + \Gamma (h_j \to c c) + \Gamma (h_j \to s s) \cr
& &\mbox{} + \Gamma (h_j \to gg)
+ \Gamma (h_j \to WW) + \Gamma (h_j \to ZZ) +  \Gamma (h_j \to t t)  \ ,
\end{eqnarray}
where the notations for each particle can be understood without difficulty.

Note that the Higgs bosons might decay into a supersymmetric particle pair if their masses are small enough.
For a comprehensive analysis, the Higgs decays into sfermions, neutralinos, and chargions also must be considered
in explicit CP violation scenario on the present model.
The Higgs decays into a superparticle pair do not play an important role for relatively light Higgs bosons [30].
Also, it has been noted that the decay modes into supersymmetric particles acquire significant branching ratios
and can be dominant ones if they are possible [30].
In our case, we set the parameter values such that a common SUSY breaking scale is taken to be 1 TeV.
Thus, the squark masses are relatively heavy in our case, and therefore we speculate
that the Higgs decays into a squark pair would contribute weakly to the total Higgs decay.
We note that our speculation is based on the size of the coupling coefficients
between Higgs bosons and a pair of squarks, which we explicitly derive.
Using them, we calculate the partial decay width of a Higgs boson into a pair of gluons via squarks.
It remains as a future study to calculate the Higgs decays into a pair of supersymmetric particles
in order to examine our speculations on its relative dominance in the total Higgs decay.

The partial decay width of a neutral Higgs boson $h_j$ into a pair of down or up quarks is given as
\begin{eqnarray}
\Gamma (h_j \to d {\bar d}) & = & {C_f g_2^2 m_d^2 m_{h_j} \over 32 \pi m_W^2}
\sqrt{1- {4 m_d^2 \over m_{h_j}^2}}
\left[ (G^S_{h_j dd})^2 \left (1 - {4 m_d^2 \over m_{h_j}^2} \right )
+ (G^P_{h_j dd})^2 \right ] \ , \cr
\Gamma (h_j \to u {\bar u}) & = & {C_f g_2^2 m_u^2 m_{h_j } \over 32 \pi m_W^2}
\sqrt{1- {4 m_u^2 \over m_{h_j }^2}}
\left[ (G^S_{h_j uu})^2 \left (1 - {4 m_u^2 \over m_{h_j }^2} \right )
+ (G^P_{h_j uu})^2 \right ] \ ,
\end{eqnarray}
where
\begin{eqnarray}
&& G^S_{h_j dd} = {O_{1j} \over \cos \beta}   \ , \quad
G^S_{h_j uu} = {O_{2j} \over \sin \beta} \ ,     \cr
& & G^P_{h_j dd} = \tan \beta O_{3j} \ , \quad
G^P_{h_j uu} = \cot \beta O_{3j} \ ,
\end{eqnarray}
where $O_{ij}$ are elements of the orthogonal
transformation matrix which diagonalizes the mass matrix for the five neutral Higgs bosons.
The color factor is  $C_f= 3$ for quarks and $C_f= 1$ for leptons.
The partial decay width of a neutral Higgs boson $h_j$ into a pair of charged leptons
is the same as $\Gamma (h_j \to d {\bar d})$.

The partial decay width of $h_j$ into a pair of weak gauge bosons may be obtained from that
of the SM Higgs boson, through a relation given by
\begin{equation}
\Gamma (h_j \to VV) = G_{h_j VV}^2 \ \Gamma_{\rm SM} (h_j \to VV) \ ,
\end{equation}
where $\Gamma_{\rm SM} (h_l \to VV)$ is the decay width of the SM Higgs boson into a pair of gauge bosons,
one of produced gauge bosons being virtual,
\begin{equation}
G_{h_j VV} = \cos \beta O_{1j} + \sin \beta O_{2j}  \ .
\end{equation}
The partial decay width of $h_j$ into a gluon pair is given as
\begin{equation}
\Gamma (h_j \to gg) = {\alpha_s^2 (m_Z) m_{h_j}^3 \over 64 \pi^3 v^2} ( |A^S|^2 + |A^P|^2 ) \ ,
\end{equation}
where $\alpha_s (m_Z)$ is the coupling coefficient of the strong interactions, evaluated at the electroweak scale,
and $A^S$ and $A^P$ are respectively the scalar and pseudoscalar gluon amplitudes, which are given as
\begin{eqnarray}
A^S & = & \sum_{q=t,b} \left[
G^S_{h_j qq} A_q^S (\tau_q)
+ \sum_{a=1,2} G_{h_j{\tilde q}_a {\tilde q}_a }
{v^2 \over 2 m_{{\tilde q}_a}^2} A^S_{\tilde q} (\tau_{\tilde q})
 \right ] \ , \cr
A^P & = & \sum_{q=t,b} \left[ G^P_{h_j qq} A_q^P (\tau_q) \right ] \ ,
\end{eqnarray}
where $G_{h_j {\tilde q}_a {\tilde q}_a}$ is the coupling coefficient of the neutral Higgs bosons to a squark pair,
$A^S_q (\tau_q)$ and $A^P_q (\tau_q)$ are respectively the scalar and pseudoscalar form factors due to the quark,
and $A^S_{\tilde q} (\tau_{\tilde q})$ is the scalar form factor due to the squark, given as
\begin{equation}
A^S_q (\tau_q) = \tau_q [1 + (1 - \tau_q) f(\tau_q) ] \ , \quad
A^P_q (\tau_q) = \tau_q f(\tau_q) \ , \quad
A^S_{\tilde q} (\tau_{\tilde q}) = \tau_{\tilde q} [ \tau_{\tilde q} f(\tau_{\tilde q}) - 1] \ ,
\end{equation}
with the scaled variables defined as
\begin{equation}
\tau_q = {4 m_{q_b}^2 \over m_{h_j}^2 } \ , \quad
\tau_{\tilde q} = {4 m_{{\tilde q}_b}^2 \over m_{h_j}^2 }   \ ,
\end{equation}
and the function $f$ defined as
\begin{equation}
f(\tau) = \left \{
\begin{array}{cl}
{\rm arcsin}^2(1/\sqrt{\tau})    & \qquad \tau \geq 1 \ , \cr
-{1 \over 4} \left[ \log  \left( \displaystyle{{1+\sqrt{1+ \tau} \over 1 - \sqrt{1- \tau}} } \right) -i \pi \right]^2
& \qquad \tau < 1 \ .
\end{array} \right .
\end{equation}

Notice that there is squark contributions to the partial decay width of $h_j$
due to the couplings of the neutral Higgs boson to a pair of squarks of the third generation.
Let us describe the couplings of the neutral Higgs bosons to a squark pair in more detail.
They are given as
\begin{equation}
G_{h_j {\tilde q}_l {\tilde q}_m}
= \left ( \Gamma^{h_i {\tilde q} {\tilde q} }   \right )_{rs}
    O_{ij} U_{rl}^{{\tilde q} *} U_{sm}^{\tilde q} \ ,
\end{equation}
where $U^{\tilde q}$  is the unitary matrix that transforms the weak eigenstates of
the squark ${\tilde q}$ ($q = t, b$) to the mass eigenstates,
$\Gamma^{h_i {\tilde q} {\tilde q} }$ is the coupling of $h_j$ to a pair of the weak eigenstates of
the squark ${\tilde q}$, and the subscript indices run $l,m =1,2$ and $r,s$ stand for $L,R$.
We may parameterize $U^{\tilde q}$ as
\begin{equation}
U^{\tilde q} =
\left(\begin{array}{cc}
\cos \theta_{\tilde q} & - \sin \theta_{\tilde q} e^{-i \phi_{\tilde q}} \\
\sin \theta_{\tilde q} e^{i \phi_q} & \cos \theta_{\tilde q}  \
\end{array} \right)  \ ,
\end{equation}
where the mixing angles $\theta_{\tilde q}$ and the complex phases $\phi_{\tilde q}$ vary
between $-\pi/2$ and $\pi/2$, and thus both $\cos \theta_{\tilde q}$ and $\cos \phi_{\tilde q}$ are non-negative.
We obtain the explicit expressions for $\Gamma^{h_i {\tilde q} {\tilde q} }$ ($q = t, b$) as follows:
\begin{eqnarray}
\Gamma^{h_1 {\tilde t} {\tilde t} } & = & {m_t \over v^2 \sin \beta}
\left(\begin{array}{cc}
0 & \lambda x  \\
\lambda x & 0  \
\end{array} \right)  \ , \cr
\Gamma^{h_2 {\tilde t} {\tilde t} } & = & {m_t \over v^2 \sin \beta}
\left(\begin{array}{cc}
2 m_t & - A_t e^{-i \phi_t}  \\
- A_t e^{+ i \phi_t} & 2 m_t  \
\end{array} \right)  \ , \cr
\Gamma^{h_3 {\tilde t} {\tilde t} } & = & {m_t \over v \sin \beta}
\left(\begin{array}{cc}
0 & \lambda \cos \beta  \\
\lambda \cos \beta & 0  \
\end{array} \right)  \ , \cr
\Gamma^{h_4 {\tilde t} {\tilde t} } & = & {m_t \over v^2}
\left(\begin{array}{cc}
0 & i (A_t e^{-i \phi_t} \cot \beta + \lambda x)  \\
 -i (A_t e^{+ i \phi_t} \cot \beta + \lambda x) & 0  \
\end{array} \right)  \ , \cr
\Gamma^{h_5 {\tilde t} {\tilde t} } & = & {m_t \over v}
\left(\begin{array}{cc}
0 & i \lambda \cot \beta  \\
- i \lambda \cot \beta & 0  \
\end{array} \right)  \ , \cr
\Gamma^{h_1 {\tilde b} {\tilde b} } & = & {m_b \over v^2 \cos \beta}
\left(\begin{array}{cc}
2 m_b & - A_b e^{- i \phi_b} \\
- A_b e^{+ i \phi_b} & 2 m_b  \
\end{array} \right)  \ , \cr
\Gamma^{h_2 {\tilde b} {\tilde b} } & = & {m_b \over v^2 \cos \beta}
\left(\begin{array}{cc}
0 & \lambda x  \\
\lambda x & 0  \
\end{array} \right)  \ , \cr
\Gamma^{h_3 {\tilde b} {\tilde b} } & = & {m_b \over v \cos \beta}
\left(\begin{array}{cc}
0 & \lambda \sin \beta  \\
\lambda \sin \beta & 0  \
\end{array} \right)  \ ,  \cr
\Gamma^{h_4 {\tilde b} {\tilde b} } & = & {m_b \over v^2}
\left(\begin{array}{cc}
0 & i (A_b e^{-i \phi_b} \tan \beta + \lambda x)  \\
 -i (A_b e^{+ i \phi_b} \tan \beta + \lambda x) & 0  \
\end{array} \right)  \ , \cr
\Gamma^{h_5 {\tilde b} {\tilde b} } & = & {m_b \over v}
\left(\begin{array}{cc}
0 & i \lambda \tan \beta  \\
- i \lambda \tan \beta & 0  \
\end{array} \right)  \ .
\end{eqnarray}
The above formula for the coupling coefficient between a Higgs boson and a pair of squarks may be used
for the calculation of the partial decay width of the Higgs boson into a pair of stop or sbottom quarks.

\section{NUMERICAL ANALYSIS}

For our numerical analysis, we fix some parameter values.
The renormalization scale is taken as $\Lambda = 300$ GeV.
The quark masses are taken as $m_t = 175$ GeV for top quark and $m_b = 4$ GeV for bottom quark.
The strong coupling constant is evaluated at the electroweak scale as $\alpha(m_Z) = 0.1187$,
and the weak mixing angle is set  as $\sin^2 \theta_W = 0.23$.
Then, there are a few free parameters in our model, the MNMSSM with explicit CP violation
at the one-loop level.
The CP symmetry in our model is explicitly violated at the one-loop level by the presence of $\phi_t$ and $\phi_b$,
arising from the squark contributions of the third generation in the effective potential.
Other relevant free parameters besides $\phi_t$ and $\phi_b$ are: $\tan \beta$,
$\lambda$, $A_{\lambda}$, $x$, $\xi$, $m_Q$, $A_t$, and $A_b$.
We take to be $A_t=A_b$, and $\phi_t=\phi_b$ for simplicity.

We first calculate $m_{h_1}$, the mass of the lightest neutral Higgs boson in our model,
for $m_Q = A_t=1000$ GeV, $\tan \beta =5$, $\lambda = 0.1$, $A_{\lambda} = x = 100$ GeV.
The result is shown in Fig. 1, where contours of $m_{h_1}$ is plotted on the ($\xi, \phi_t$)-plane.
Please notice that $m_{h_1}$ increases as the tadpole coefficient $\xi$ increases,
because the tadpole term prohibits a massless neutral Higgs boson.
The ($\xi, \phi_t$)-plane in Fig.1 has some shaded regions.
The dotted region is where the spontaneous symmetry breaking does not work, and the hatched regions
are experimentally excluded by the LEP2 constraint.
Only the remaining bright region is physically allowed.
Therefore, for the chosen parameter values, we find that $12 \le m_{h_1} \le 52$ GeV.

The hatched regions, imposed by the LEP2 data, may need some explanations.
We assume that the SM Higgs boson decays exclusively into a pair of bottom quarks.
Then, we apply the LEP2 data to obtain the corresponding coupling of $h_1$ in our model.
We interpret the experimental constraint upon the SM Higgs boson by LEP2 for the parameter region in our model,
with the modified Higgs coupling.
In this way, the hatched regions are established.

One may note that the allowed mass range between 12 and 52 GeV for $h_1$ in our model is
significantly below 114.5 GeV, which is the LEP2 lower bound on the SM Higgs boson.
However, it does not imply in a straightforward way that the present model is phenomenologically contradicting
the experimental constraint of LEP2.
In the CP-violating case, the mass of the neutral Higgs boson may be released from the LEP2 constraint.
For example, the MSSM in the explicit CP violation scenario with $\phi_t= \pi/2$ allows
the existence of a neutral Higgs boson with a mass as small as 30 GeV [31].
This implies that discovering a neutral Higgs boson depends not only on its mass
but also its coupling coefficients and other factors.
Since the lightest neutral Higgs boson in our model is a mixture of the neutral components of the three Higgs fields,
the possibility of its discovery would also be affected by how much the Higgs singlet field is mixed in it.

We next calculate $m_{h_i}$ ($i$ = 1-5), the masses of the five neutral Higgs boson in our model,
for $m_Q = A_t=1000$ GeV, $\tan \beta =5$, $\lambda = 0.1$, $A_{\lambda} = 100$ GeV, $x = 200$ GeV,
and $\xi = 100$ GeV.
The result is shown in Fig. 2, where $m_{h_i}$ ($i$ = 1-5) are plotted as functions of $0 \le \phi_t \le \pi$.
It can easily be observed that, for the chosen parameter values,
all of the five neutral Higgs bosons have small masses, below the top mass.
Also, the dependence of $m_{h_i}$ ($i$ = 1-5) on the CP phase $\phi_t$ is quite recognizable in Fig. 2.
As the CP phase increases from zero to $\pi$, the figure shows that $m_{h_1}$ increases from 60 GeV to 66 GeV,
$m_{h_2}$ from 68 GeV to 69 GeV,
$m_{h_3}$ from 73 GeV to 94 GeV,
$m_{h_4}$ from 79 GeV to 127 GeV, and $m_{h_5}$ from 143 GeV to 163 GeV.
Thus, the neutral Higgs bosons deviate 10 \%, 1.4 \%, 30 \%, 60 \%, and 14 \% in their masses, respectively,
for $0 \le \phi_t \le \pi$.
The variation of the neutral Higgs boson masses in the MNMSSM against the CP phase may be regarded to be very large
as compared to the result in the NMSSM [16], where the five neutral Higgs boson masses vary
below 5 \% against the CP phase.

We continue our numerical analysis to calculate the branching ratios of each of the five neutral Higgs bosons.
For the same parameter values as in Fig. 2, we obtain a number of important branching ratios.
The results are shown in Figs. 3a-3e, where they are plotted  as functions of $0 \le \phi_t \le \pi$.
Note that for the whole range of the CP phase the decay channel into a pair of bottom quarks is
most dominant for all five neutral Higgs bosons.
The branching ratio of $BR (h_i \to b{\bar b})$ is invariably more than 90 \% for all of $h_i$ ($i$ = 1-5).

For the branching ratios of $h_5$, the heaviest neutral Higgs boson in our model,
are somewhat different from the other four neutral Higgs bosons.
Since its mass is calculated to be between 143 GeV and 163 GeV for $0 \le \phi_t \le \pi$,
its decay into a pair of weak gauge boson might be important, as Fig. 3e indicates.
It is noticeable that $BR (h_5 \to WW)$  is comparable to $BR (h_5 \to b {\bar b})$ for $\phi_t \sim \pi$.
The fluctuations of the branching ratios of $h_5$ against the CP phase are also recognizable in Fig. 3e.
For $h_5$ decay, $BR(h_5 \to \tau^+ \tau^-)$ and $BR (h_5 \to ZZ)$ remains almost stable,
whereas  $BR(h_5 \to c{\bar c})$ and $BR (h_5 \to gg)$ decrease by an order of magnitude,
as $\phi_t$ varies from zero to $\pi$.

The patterns of the branching ratios of $h_1$, $h_2$, $h_3$, and $h_4$, are nearly similar to each other.
The next dominant decay channel is $h_i \to \tau^+ \tau^-$ for $i$ = 1-4.
The branching ratios do not wildly fluctuate against the CP phase, and the relative size
between the branching ratios are
$BR(h_i \to  b {\bar b}) >BR(h_i \to \tau^+ \tau^-) > BR(h_i \to c {\bar c}) > BR(h_i \to gg)$, for $i$ = 1-4.

We would like to remark the interesting behavior of $h_4$ at the branching ratio
for the gluon channel $BR(h_4 \to gg)$.
As shown in Fig. 3d, we have $BR(h_4 \to gg) = 0.9852 \times 10^{-8}$ for $\phi_t \sim 0.9331$.
We note that the most part of this value comes from the scalar gluon amplitude $A^S$
rather than the pseudoscalar gluon amplitude $A^P$
for $\phi_t = 0.9331$.
This is  because the top and bottom quark contributions cancel out accidently each other
in the pseudoscalar amplitude $A^P$ for $h_4$ in Eq. (20),
That is, $G^P_{h_4 tt} A_t^P (\tau_t)$ and $G^P_{h_4 bb} A_b^P (\tau_b)$ in $A^P$ have
accidentally almost the same value but with  opposite sign, for $\phi_t = 0.9331$.
Thus, the scalar gluon amplitude contributes dominantly to the $BR(h_4 \to gg)$ in our analysis.

Finally, we calculate $\Gamma (h_i)$, the total decay widths of $h_i$ ($i$ = 1-5),
for the same parameter values as in Fig. 2.
The result is shown in Fig. 4, where $\Gamma(h_i)$ ($i$ = 1-5) are plotted as functions of $0 \le \phi_t \le \pi$.
Their dependence on the CP phase is quite large.
While $\Gamma(h_1)$ and $\Gamma(h_2)$ decrease as $\phi_t$ increases,
the total decay widths of the other neutral Higgs bosons increase with increasing $\phi_t$.
In particular, it is worthwhile noticing the increase of $\Gamma(h_5)$.
This behavior is understandable because the decay channel into a pair of weak gauge bosons is only allowed for $h_5$.

\section{CONCLUSIONS}

We study the Higgs sector of the MNMSSM with tadpole terms.
We see that the model can accommodate explicit CP violation at the one-loop level.
As the squarks of the third generation with non-degenerate masses give rise to the radiative corrections,
the explicit CP violation can be generated by complex phases $\phi_t$ and $\phi_b$
which appear in their mass matrices.
At the tree level, neither explicit nor spontaneous CP violation would be viable in the Higgs sector of the model.
Unlike our model, the NMSSM may possess complex phases at the tree level.

In the presence of the complex phases that trigger the explicit CP violation at the one-loop level,
we calculate the masses, the branching ratios for dominant decay channels,
and the decay widths of the five neutral Higgs bosons in our model.
The masses show fluctuations up to 60 \% as the CP phase varies, for given parameter values.
Whereas the heaviest neutral Higgs boson in our model can be as heavy as 160 GeV,
the rest of them is relatively light.
In particular, the mass of the lightest neutral Higgs boson in our model is predicted to be as small as 12 GeV,
for reasonable ranges of relevant parameter values.
Although the predicted mass of the lightest neutral Higgs boson is well below experimental constraint of LEP2,
it might have been escaped from experiments since its discovery depends
on its coupling coefficients to other particles as well as on its mass.

The coupling coefficients of the neutral Higgs bosons to fermion pairs, weak gauge boson pairs,
gluon pairs, and squark pairs have been explicitly calculated.
The results of calculation, which have not been derived before,
may be used helpfully for the search of Higgs boson in similar supersymmetric models in the future experiments.
In terms of these coupling coefficients, the branching ratios of the five neutral Higgs bosons
for dominant decay channels are calculated, where the CP violation effect is included.
We find that the heaviest neutral Higgs boson exhibits fluctuating branching ratios
against the variation of the CP phase.
For the rest four Higgs bosons, the decay into a pair of bottom quarks is most dominant,
and the branching ratios are rather stable against the variation of the CP phase.
Finally, we calculate the total decay widths of the five neutral Higgs bosons,
which are significantly dependent on the CP phase.

If CP is violated at the one-loop level in the MNMSSM with tadpole terms,
the result of our calculations, which has been done for a representative set or range of parameter values,
suggests that the whole parameter space should be investigated
in order to draw more interesting observations within the context of CP violation.

\vskip 0.3 in
\noindent
{\large {\bf ACKNOWLEDGMENTS}}
\vskip 0.2 in
S. W. Ham thanks John F. Gunion, Jihn E. Kim, and Radovan Dermisek for valuable comments.
S. W. Ham is partly supported by MEST in 2007 (No. K2071200000107 A020000110) and
in part by the Korea Research Foundation Grant funded by the Korean Government
(MOEHRD, Basic Research Promotion Fund) (KRF-2007-000-C00010).
This work is supported by Konkuk University in 2007.




\vfil\eject
{\noindent\bf FIGURE CAPTIONS}

\vskip 0.25 in
\noindent
Fig. 1: Contours of the lightest neutral Higgs boson mass on ($\xi, \phi_t$)-plane,
for $m_Q = A_t=1000$ GeV, $\tan \beta =5$, $\lambda = 0.1$, and $A_{\lambda} = x = 100$ GeV.
The bright region is physically allowed whereas the shaded regions are excluded.
The spontaneous symmetry breaking does not work in the dotted region,
and the experimental constraint by the LEP2 data excludes the hatched regions.

\vskip 0.25 in
\noindent
Fig. 2: The masses of the five neutral Higgs bosons as functions of $\phi_t$,
where the parameters are set as $m_Q = A_t=1000$ GeV, $\tan \beta =5$, $\lambda = 0.1$,
$A_{\lambda} = x/2 = \xi = 100$ GeV.

\vskip 0.25 in
\noindent
Fig. 3(a): The branching ratios of $h_1$ as functions of $\phi_t$, for the same parameter values as Fig. 2.

\vskip 0.25 in
\noindent
Fig. 3(b): The branching ratios of $h_2$ as functions of $\phi_t$, for the same parameter values as Fig. 2.

\vskip 0.25 in
\noindent
Fig. 3(c): The branching ratios of $h_3$ as functions of $\phi_t$, for the same parameter values as Fig. 2.

\vskip 0.25 in
\noindent
Fig. 3(d): The branching ratios of $h_4$ as functions of $\phi_t$, for the same parameter values as Fig. 2.

\vskip 0.25 in
\noindent
Fig. 3(e): The branching ratios of $h_5$ as functions of $\phi_t$, for the same parameter values as Fig. 2.

\vskip 0.25 in
\noindent
Fig. 4: The total decay widths of the five neutral Higgs bosons as functions of $\phi_t$,
for the same parameter values as Fig. 2.

\vfil\eject

\renewcommand\thefigure{1}
\begin{figure}[t]
\begin{center}
\includegraphics[scale=0.6]{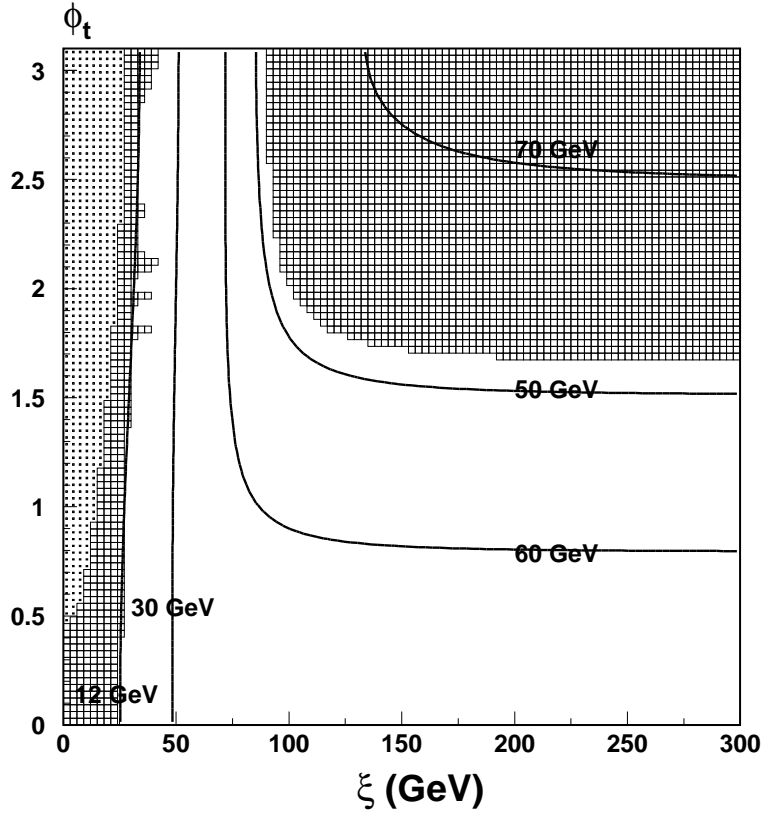}
\caption[plot]{Contours of the lightest neutral Higgs boson mass on ($\xi, \phi_t$)-plane,
for $m_Q = A_t=1000$ GeV, $\tan \beta =5$, $\lambda = 0.1$, and $A_{\lambda} = x = 100$ GeV.
The bright region is physically allowed whereas the shaded regions are excluded.
The spontaneous symmetry breaking does not work in the dotted region,
and the experimental constraint by the LEP2 data excludes the hatched regions.}
\end{center}
\end{figure}

\renewcommand\thefigure{2}
\begin{figure}[t]
\begin{center}
\includegraphics[scale=0.6]{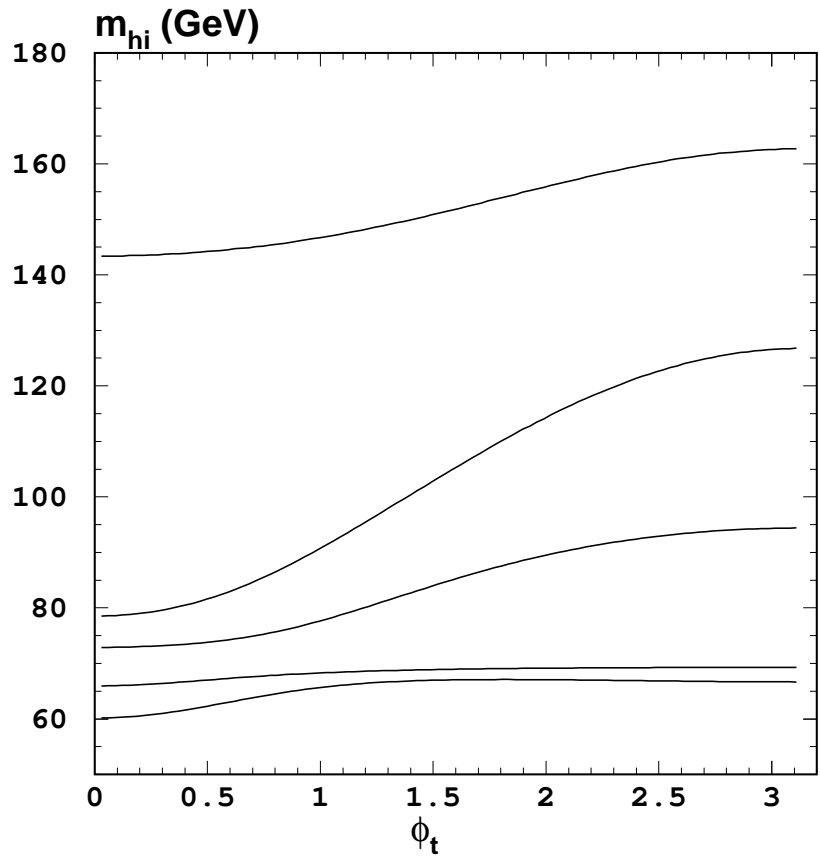}
\caption[plot]{The masses of the five neutral Higgs bosons as functions of $\phi_t$,
where the parameters are set as $m_Q = A_t=1000$ GeV, $\tan \beta =5$, $\lambda = 0.1$,
$A_{\lambda} = x/2 = \xi = 100$ GeV.}
\end{center}
\end{figure}

\renewcommand\thefigure{3a}
\begin{figure}[t]
\begin{center}
\includegraphics[scale=0.6]{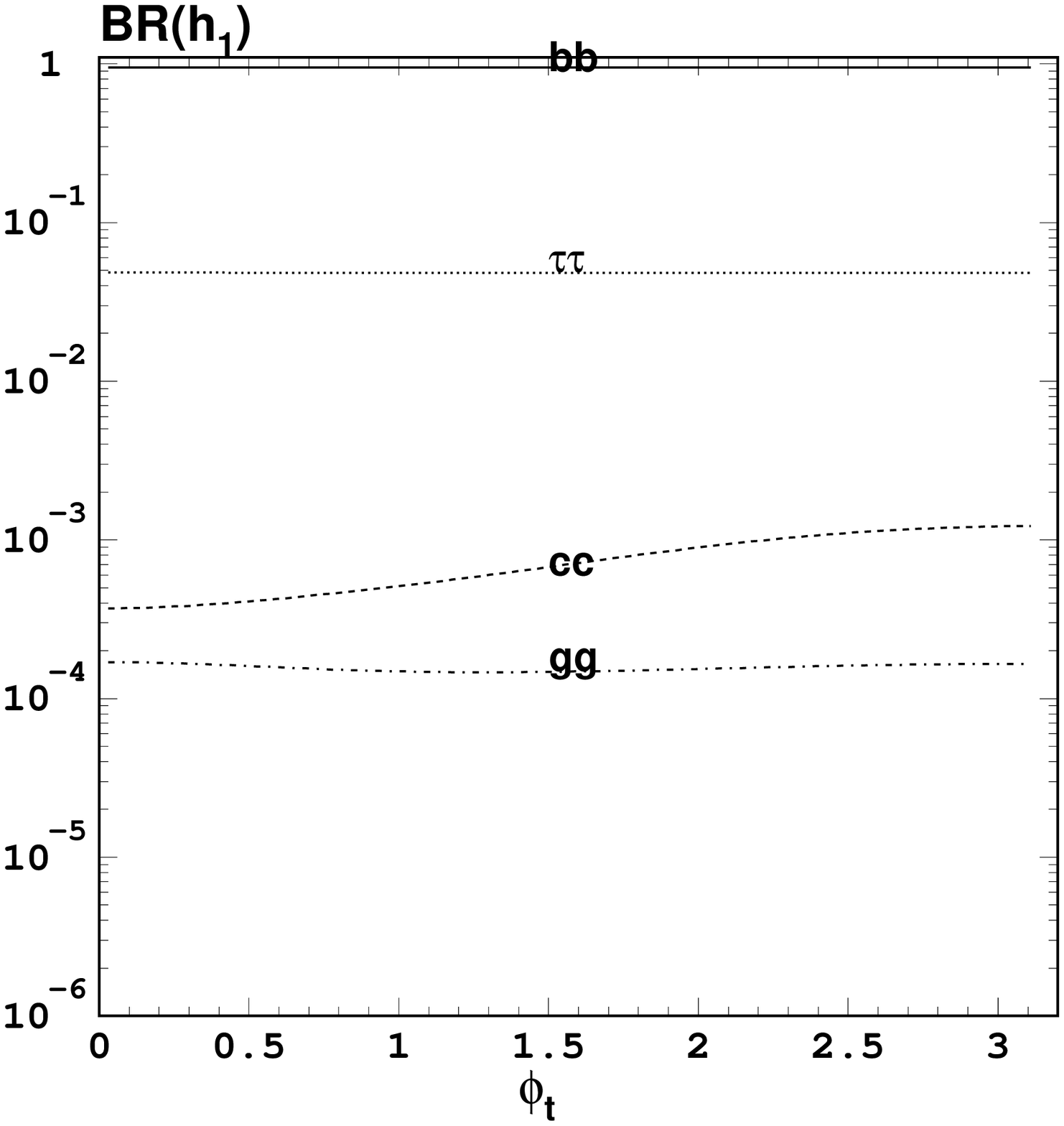}
\caption[plot]{The branching ratios of $h_1$ as functions of $\phi_t$, for the same parameters as Fig. 2.}
\end{center}
\end{figure}

\renewcommand\thefigure{3b}
\begin{figure}[t]
\begin{center}
\includegraphics[scale=0.6]{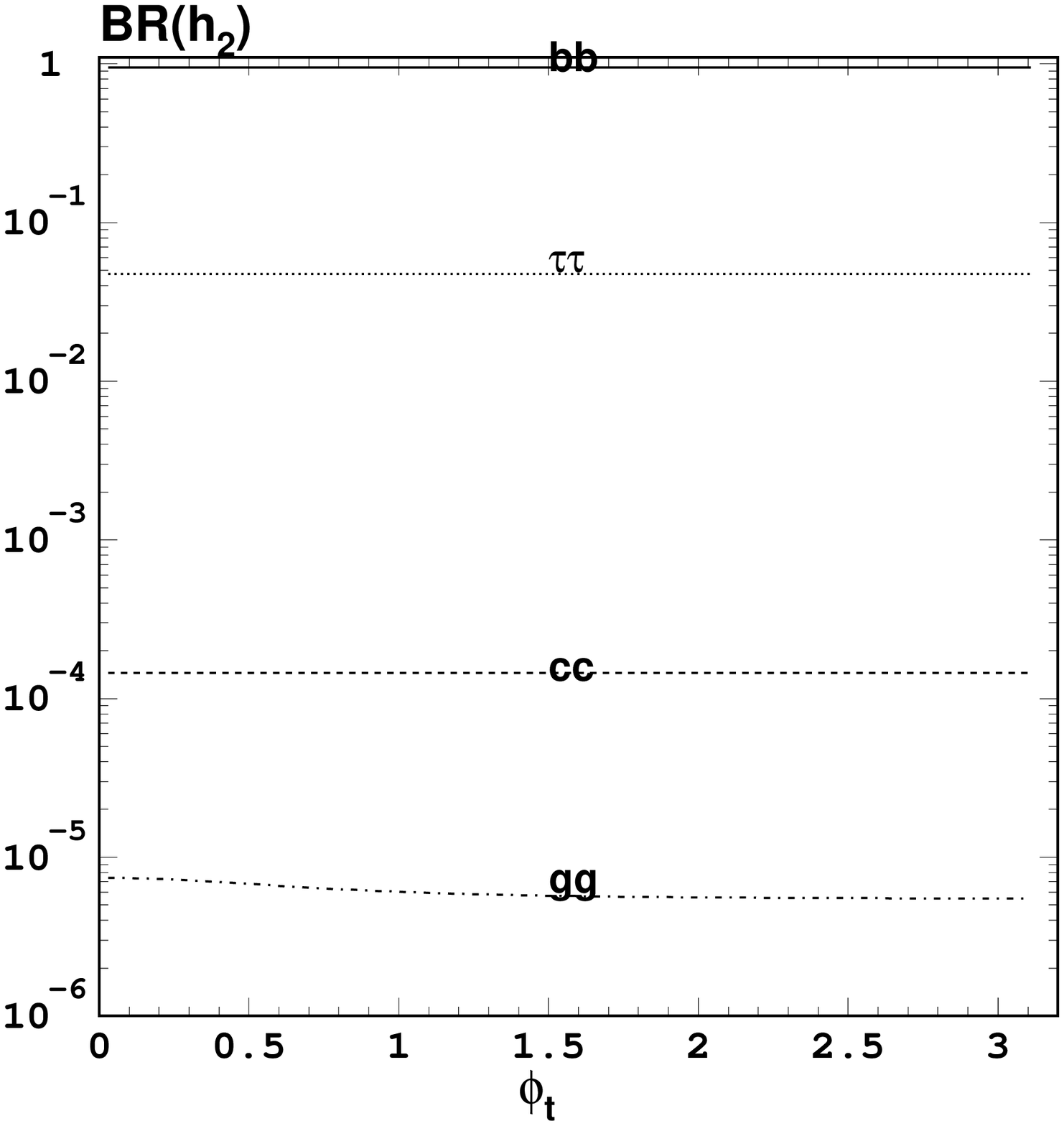}
\caption[plot]{The branching ratios of $h_2$ as functions of $\phi_t$, for the same parameters as Fig. 2.}
\end{center}
\end{figure}

\renewcommand\thefigure{3c}
\begin{figure}[t]
\begin{center}
\includegraphics[scale=0.6]{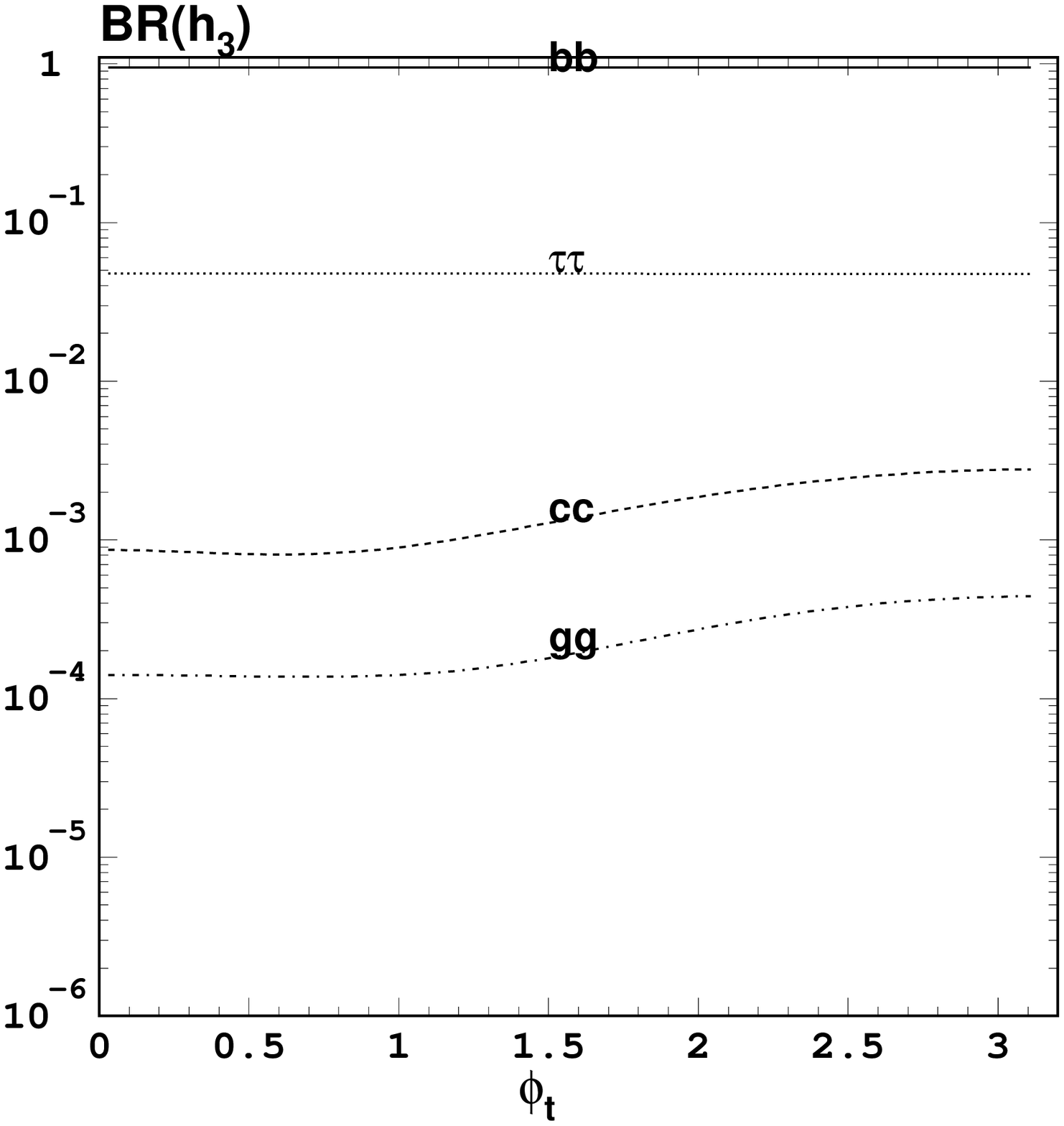}
\caption[plot]{The branching ratios of $h_3$ as functions of $\phi_t$, for the same parameters as Fig. 2.}
\end{center}
\end{figure}

\renewcommand\thefigure{3d}
\begin{figure}[t]
\begin{center}
\includegraphics[scale=0.6]{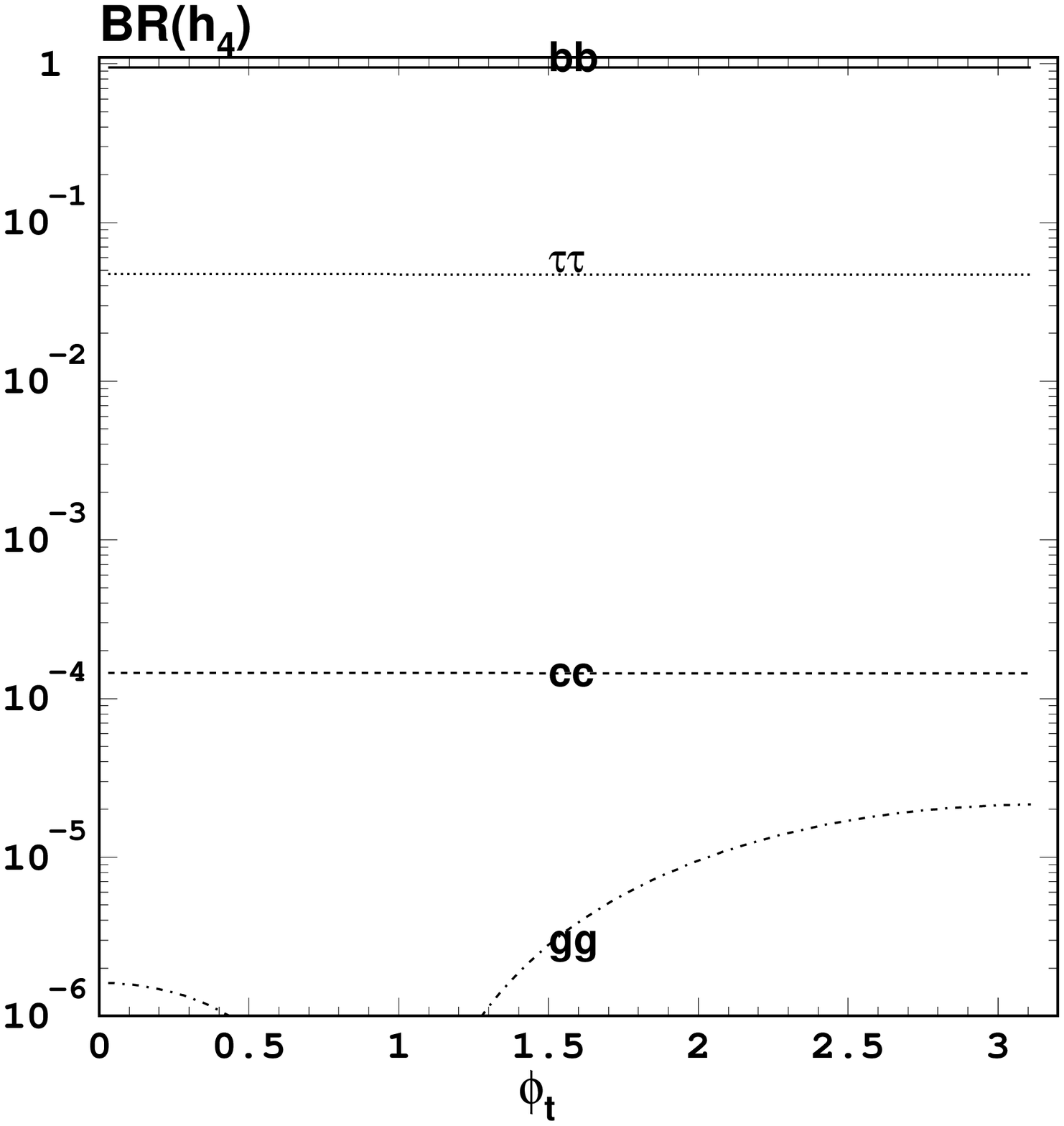}
\caption[plot]{The branching ratios of $h_4$ as functions of $\phi_t$, for the same parameters as Fig. 2.}

\end{center}
\end{figure}

\renewcommand\thefigure{3e}
\begin{figure}[t]
\begin{center}
\includegraphics[scale=0.6]{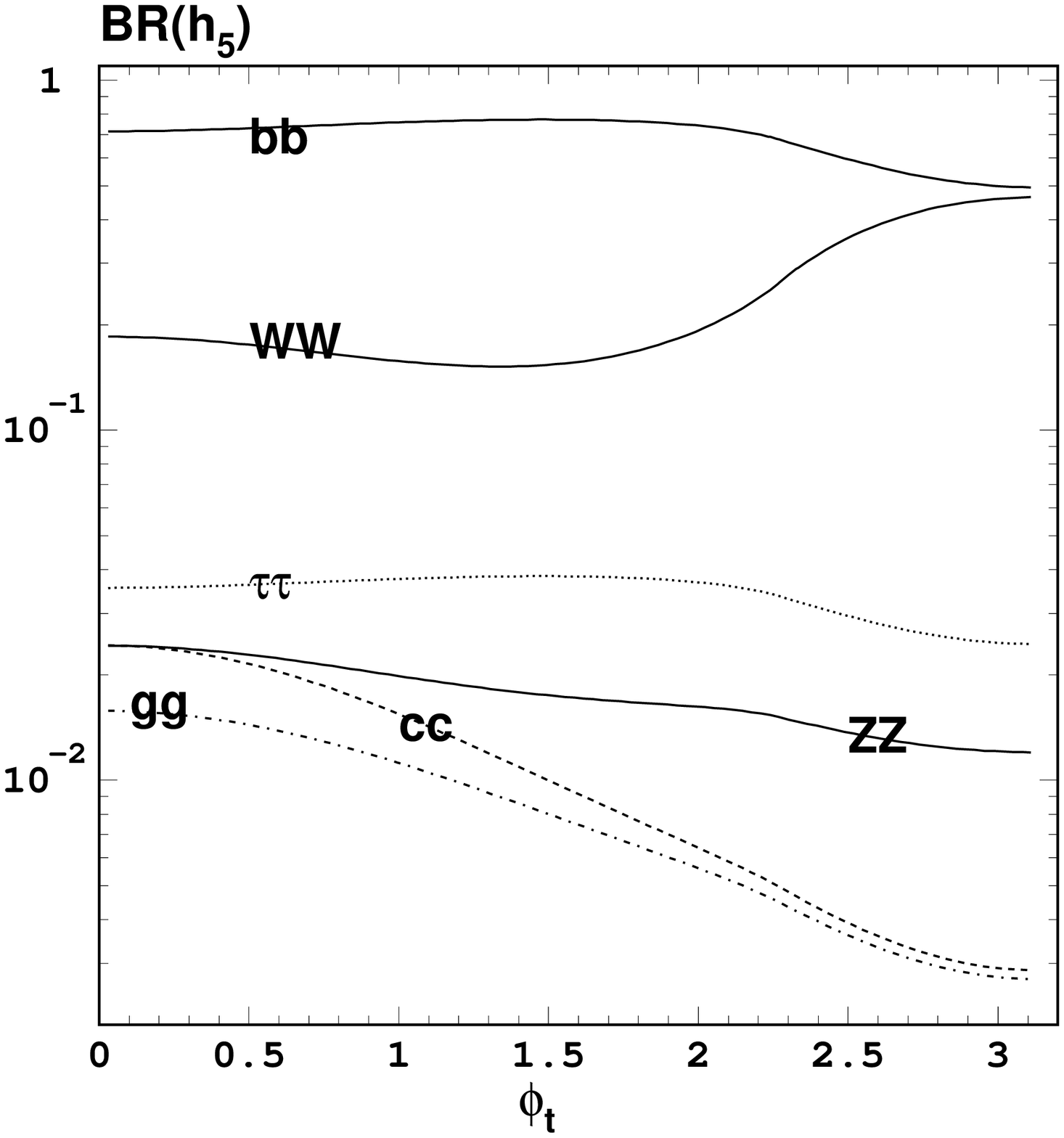}
\caption[plot]{The branching ratios of $h_5$ as functions of $\phi_t$, for the same parameters as Fig. 2.}
\end{center}
\end{figure}

\renewcommand\thefigure{4}
\begin{figure}[t]
\begin{center}
\includegraphics[scale=0.6]{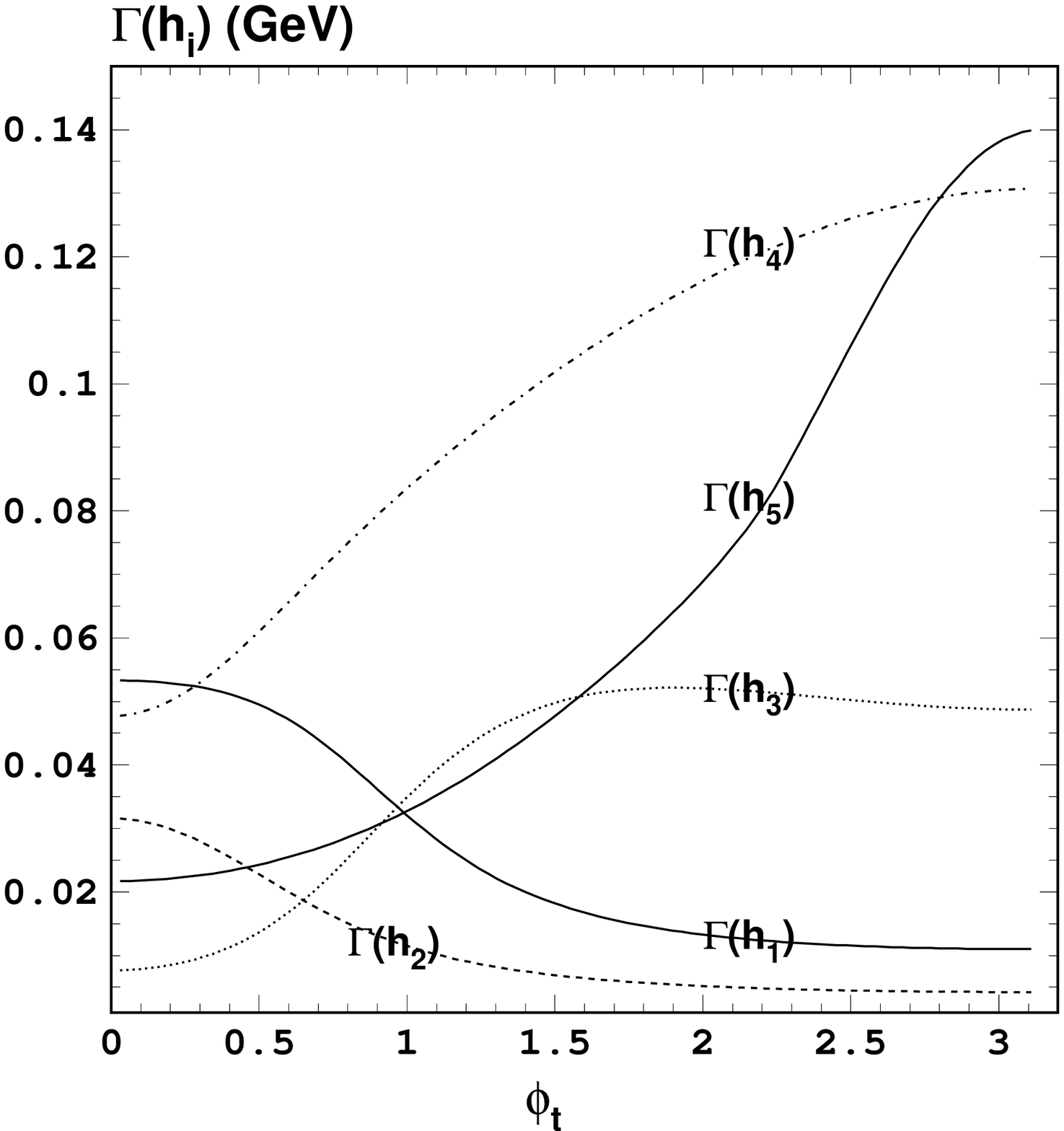}
\caption[plot]{The total decay widths of the five neutral Higgs bosons as functions of $\phi_t$,
for the same parameters as Fig. 2.
}
\end{center}
\end{figure}

\end{document}